\documentclass[prints]{aa}
\usepackage{graphicx}
\def\logz{\lbrack\hbox{Fe/H}\rbrack}
\begin{document}
\title{A new galaxy near the Local Group in Draco
\thanks{Based on observations made with the NASA/ESA Hubble Space
Telescope.  The Space Telescope Science Institute is operated by the
Association of Universities for Research in Astronomy, Inc. under NASA
contract NAS 5--26555. Based in part on observations obtained with
the 6-m telescope operated by the Russian Academy of Sciences}}
\titlerunning{New dwarf galaxy in Draco}
\author{I.D.Karachentsev \inst{1} \and M.E.Sharina \inst{1}
\and A.E.Dolphin \inst{2}
\and D.Geisler \inst{3}
\and E.K.Grebel \inst{4}
\and P.Guhathakurta \inst{5}\thanks{Alfred P.\ Sloan Research Fellow}
\and P.W.Hodge \inst{6}
\and V.E.Karachentseva \inst{7}
\and A.Sarajedini \inst{8}
\and P.Seitzer \inst{9}}
\authorrunning{I.D.Karachentsev et al.}
\institute{Special Astrophysical Observatory, Russian Academy
of Sciences, N.Arkhyz, KChR, 369167, Russia,
\and Kitt Peak National Observatory, National Optical Astronomy Observatories,
P.O. Box 26732, Tucson, AZ 85726, USA
\and Departamento de Fisica, Grupo de Astronomia, Universidad de Concepcion,
Casilla 160-C, Concepcion, Chile
\and Max-Planck-Institut f\"{u}r Astronomie, K\"{o}nigstuhl 17, D-69117
Heidelberg, Germany
\and UCO/Lick Observatory, University of California at Santa Cruz, Santa Cruz,
CA 95064, USA
\and Department of Astronomy, University of Washington, Box 351580, Seattle, WA
\and Astronomical observatory of Kiev University, 04053, Observatorna 3, Kiev,
Ukraine
\and Department of Astronomy, University of Florida, Gainesville, FL 32611, USA
\and Department of Astronomy, University of Michigan, 830 Dennison Building,
Ann Arbor, MI 48109, USA}
\date{Received: 31 July 2001/Accepted 12 September 2001}
\abstract{
We present HST WFPC2 and ground-based images of the low surface brightness
dwarf $Irr/Sph$ galaxy KKR~25 in Draco. Its colour-magnitude diagram shows
red giant branch stars with the tip at $I= 22\fm32$, and the presence of
some blue stars. The derived true distance modulus, $26\fm35 \pm 0\fm14$,
corresponds to linear distances of KKR~25 from the Milky Way and from the
Local Group centroid of 1.86 and 1.79 Mpc, respectively. The absolute magnitude
of the galaxy, $M_V = -10.48$, its linear diameter ( 0.54 Kpc) and central
surface brightness ($\Sigma_{0,V}  = 24\fm0\pm0\fm2/\sq\arcsec $ ) are
typical of other $dIrr/dSphs$ in the Local Group. Being situated just
beyond the radius of the zero-velocity surface of the Local Group, KKR~25
moves away from the LG centroid at a velocity of $V_{LG} = + 72 $ km/s.
\keywords{ galaxies: dwarf --- galaxies:
distances }}
\maketitle

\section{Introduction}

  Among the faintest dwarf irregular ( dIrr ) galaxies in the Local Group,
there are galaxies that may be in transition from dIrrs to dwarf
spheroidals ( dSphs ): Phoenix ( $M_V=-9.8$ ) and LGS 3 ( $M_V=-10.5$ ) .
These galaxies contain prominent
old populations reminiscent of dSphs, but also some recent star formation
as indicated by young blue stars and have HI associated with
them, which is characteristic for dIrrs. They are therefore classified
as dIrr/dSphs (Grebel 2000). The discovery of Antlia dwarf (Whiting et
al. 1997) with $M_V= -10.7$  at a distance of $D$ = 1.32 Mpc,
shows that such extremely faint bluish galaxies
of regular shape may also occur outside the Local Group. A detailed
studying of these tiny transient objects is very important to
understand their origin and evolution.

  The detection of low surface brightness galaxies with an
absolute magnitude of about $-10^m$, situated between galaxy groups, is a
complicated observational task. Recent all-sky searches for nearby dwarf
galaxies made on the POSS-II and ESO/SERC plates by Karachentseva
\& Karachentsev (1998), Karachentseva et al. (1999,2000), and
Karachentsev et al. (2000) has led to the discovery of $\sim$600 objects
mainly of low surface brightness. One object from these lists, KKR~25,
(Karachentseva et al. 1999) is considered in the present Note.

KKR~25 has an apparent size of 1$\farcm1\times0\farcm$65  and is located next
to a luminous red foreground star. In its direction Huchtmeier et al. (2000)
detected HI emission with a heliocentric radial velocity  $V_h$ =
$(-135\pm2)$ km/s and a linewidth $W_{50} = 14$ km/s, which was
attributed to local Galactic hydrogen.
The first large-scale image of KKR 25 was obtained
with the 6-m telescope (SAO, Russia) in the $R$ and $I$ bands with a
 1$\farcs1$ seeing (FWHM) on June 18, 1999. The R image,
presented in Fig.1, shows that the galaxy is well resolved into stars.
The brightest stars in the galaxy body have a red color $(R-I)\sim 0.7$
and an apparent
magnitude  $m_I\sim22^m$. The preliminary distance estimate, mentioned by
Karachentsev \& Makarov (2001), yields for KKR~25  $D\sim1.5$ Mpc. Due to
this very close distance, we included KKR~25 in our ongoing Hubble Space
Telescope snapshot survey.

  \section { HST WFPC2 observations and data reduction}

  Observations of KKR 25 with the Hubble Space Telescope WFPC2 were obtained
on May 25, 2001 as part of the HST snapshot survey of probable nearby
galaxies (GO program 8601, PI: P.Seitzer). The galaxy was imaged in
F606W and F814W with exposure times of 600$^s$ each, with the galaxy center
located in the WF3 chip. Fig. 2 shows the galaxy image on this chip
resulting from the combination of both filters to remove cosmic rays.

  The photometric pipeline used for the snapshot survey has been described
in detail in Dolphin et al. (2001), and what follows is only a summary.
After obtaining the calibrated images from STScI, cosmic ray cleaning
was performed with the HSTphot (Dolphin 2000a) \textit{cleansep} routine
which cleans images taken with different filters by allowing for a
colour variation.  Stellar photometry was then obtained with the HSTphot
\textit{multiphot} routine which measures magnitudes simultaneously in
the two images, accounting for image alignment, WFPC2's wavelength-dependent
plate scale, and geometric distortion.
The final photometry was then done using aperture corrections to a
$0\farcs5$ radius, and the Dolphin (2000b) charge-transfer inefficiency (CTE)
correction and zero-point calibration applied. We estimate the aperture corrections
in the three wide field chips to be accurate to 0.05 magnitude. The CTE
correction depends on the X- and Y- positions, the background counts,
the brightness of the stars, and the time of the observations. For our
data the mean CTE correction makes a star brighter by $\sim0\fm25$.
Because of the small areal coverage of the Planetary Camera (PC) and
consequent lack of stars suitable for an
accurate aperture correction, the PC photometry is excluded from our
analysis. Aditionally, stars with a signal-to-noise ratio $ < 5 $,
$\mid$chi$\mid~> 2.0$, or  $\mid$sharpness$\mid~> 0.4$  in each
exposure were eliminated from the final photometry list,
in order to minimize the number of false detections.

  Finally, the F606W and F814W instrumental magnitudes
of 1875 stars were converted to
the standard $V,I$ system following the "synthetic" transformations of
Holtzman et al. (1995). We used the parameters of transformation from their
Table 10, taking into account different relations for blue and red stars
separately. Because we used the non-standard $V$ filter F606W instead of
F555W, the resulting $I$ and especially $V$ magnitudes may contain systematic
errors. However, when comparing our F606W, F814W photometry of other
snapshot targets with ground-based $V,I$ photometry, we find that the
transformation uncertainties, $\sigma {(I)}$ and  $\sigma {(V-I)}$, are
within 0$\fm$05 for stars with colours of $0 < (V-I) < 2$.  Note that
the zero-point for the F606W observations is taken from WFPC2 observations of
Omega Centauri as measured by Dolphin (2000b).

  \section{Colour-magnitude diagram and distance}

  Fig.3 shows the colour-magnitude diagrams (CMDs) derived for the
central WF3 field, as well as for the neighbouring fields WF2 and WF4.
In the central field the number of stars increases abruptly at $I\sim22\fm3$,
which we interpret as the tip of the red giant branch (TRGB). There
are also some faint, bluish stars with $(V-I) < 0.5$ in the WF3. These
stars may be indicative of a young population of KKR~25.

  The magnitude of the TRGB has been obtained applying a Sobel filter.
Following Sakai et al. (1996), we use an edge-detection filter, which is a
modified version of a Sobel kernel ([-1,0,+1]), to the Gaussian-smoothed
I-band luminosity function. Only red stars with colours 0.6 $< (V-I) <$ 1.6
were considered. The resulting luminosity function and the Sobel
filtered luminosity function are shown in Fig. 4. The TRGB corresponds
to the peak at $I$ = 22.32 (bottom panel). The peak at $I$ = 22.7 and
several others are being produced by density fluctuations along the RGB.
The error can be estimated as 1/2 of the peak width at 62\% of its maximum
and turns out to be formally 0$\fm$10.

  According to Da Costa \& Armandroff (1990), the TRGB can be assumed to
be at $M_I = -4.05$ for metal-poor systems. With a Galactic extinction
along the line of sight toward KKR~25 of $A_I = 0\fm02$  (Schlegel et al.
1998) this yields a distance modulus of $(m-M)_0 = 26\fm35\pm0\fm14$ or $D =
1.86\pm0.12$ Mpc. The quoted errors include the error in the TRGB detection
(0$\fm$10), as well as uncertainties of the HST photometry zero point
($\sim0\fm05$), the aperture corrections ($\sim0\fm05$), and crowding effects
($\sim0\fm06$) added in quadrature.

\section{Results and discussions}

  The solid line in Fig. 3 (left panel) is the M15 globular cluster
fiducial with [Fe/H] = $-$2.2 dex from Da Costa \& Armandroff (1990), which
was reddened and shifted to the derived galaxy's distance. This
low-metallicity fiducial provides a good fit to the RGB of KKR~25. With
knowledge of the distance modulus of KKR~25 we can estimate its mean
metallicity from the mean colour of the RGB measured at an absolute
magnitude $M_I = -3.5$, as recommended by Lee et al. (1993). Based on
a Gaussian fit to the colour distribution of the giant stars in the
range $22\fm6 < I < 23\fm0$  we derive a mean dereddened colour of the RGB
stars of $(V - I)_{0,-3.5}$ =  1.26$\pm$0.07. Following Lee et al. (1993)
this yields a mean metallicity  [Fe/H] = $(-2.1\pm0.3)$~dex.

  Integrated photometry of the HST data of the galaxy was 
carried out with increasing
circular apertures. The sky level was approximated by a two-dimensional
polynomial, using regions with only a few stars near the edges of the images.
Then the galaxy magnitude in each band was measured as the asymptotic
value of the derived curve of growth. Fig. 5 shows the results.

  A summary of the basic properties of KKR~25 is given in Table 1. The
parameters listed in Table 1 are: (1,2) --- equatorial coordinates of the
galaxy center, (3) --- morphological type, (4,5) --- angular diameter along
the major axis and axial ratio corresponding to a level of $B\sim26.5^m/\sq\arcsec$,
 (6) --- Galactic extinction in $V$ and $I$ bands from
Schlegel et al. (1998), (7,8) --- integrated $V$ magnitude and integrated
colour, (9) --- observed central surface brightness, (10) --- apparent $I$
magnitude of the RGB tip, (11) --- true distance modulus, (12,13) --- linear
distance from the Milky Way and from the Local Group centroid, (14,15) ---
linear diameter and absolute magnitude, (16,17) --- mean reddening-corrected
colour of the RGB tip measured at an absolute magnitude $M_I = -3.5$,
and corresponding mean metallicity,
(18) --- heliocentric velocity, (19) --- radial velocity with respect to the
Local Group centroid, (20) ---  HI line flux, (21) ---  HI line width,
(22,23) ---  hydrogen mass-to-luminosity ratio and the dynamical (virial)
mass-to-luminosity ratio determined in the same manner as in
Huchtmeier et al. (2000).

  As mentioned above, KKR~25 contains a population of faint bluish stars
with $(V-I) < 0.5$. These stars are distributed more or less homogeneously
over the galaxy body. On August 2000 KKR~25 was imaged with the 6-m telescope
with an $H_{\alpha}$  filter. After the subtraction of the continuum the
$H_{\alpha}$ image shows no compact HII regions. These properties indicate
that KKR~25 can be classified as a dIrr/dSph just as Phoenix and LGS 3.

  We searched for globular clusters in KKR~25 and found one candidate
situated half-way from the center to the left edge of WF3 (see Fig. 2).
This bright diffuse object has an integrated apparent magnitude
$V_T = 20.59$ and an angular half-light radius $r(0.5L)$ = 0$\farcs58$ that
corresponds to  $M_V = -5\fm79$ and  $R(0.5L)$ = 5.2 pc, which are within
the values typical of Galactic globular clusters. However, the integrated
color of the object, $(V-I)_T$ = 1.83, seems to be too red for a globular
cluster.

According to its distance with respect to the Local Group centroid,
$D_{LG}$ = 1.79 Mpc, KKR~25 is situated beyond the radius of the zero-velocity
surface for the Local Group (Karachentsev \& Makarov 2001). The distance
of KKR~25 is in good agreement with
its positive radial velocity $V_{LG} = +72\pm2$ km/s
with respect to the Local Group centroid. Note that the distance and the
velocity of KKR~25 are  almost the same as for another dIrr/dSph galaxy,
Antlia, which has $D_{LG}$ = 1.70 Mpc and $V_{LG}$ = +65 km/s (Aparicio
et el., 1997). Compared to Antlia, KKR~25 has a somewhat smaller luminosity
and size. But unlike Antlia (assosiated with  NGC~3109, Sex~A, and
Sex~B), KKR~25 is a totally isolated object, with no other known galaxies
within a radius of 0.5 Mpc.

\acknowledgements
{Support for this work was provided by NASA through grant GO--08601.01--A from
the Space Telescope Science Institute, which is operated by the Association
of Universities for Research in Astronomy, Inc., under NASA contract
NAS5--26555.  I.D.K., V.E.K., and E.K.G. acknowledge partial support through
the Henri Chr\'{e}tien International Research Grant administered by
the American Astronomical Society. D.G. acknowledges financial support for
this project received from CONICYT through Fondecyt grant 8000002. This
work has also been partially supported by the DFG--RFBR grant 01--02--04006
and RFBR grant 01--02--16001}

{}

\newpage
\begin{table}
\caption{Observed and derived properties of KKR 25 in Draco}
\begin{tabular}{|ll|}\hline
 Parameter                       &    KKR 25   \\  \hline
				 &                \\
 R.A. (B1950.0)                   &    16$^h$12$^m$37$\fs3$           \\
 Dec. (B1950.0)                   &   +54$\degr29\arcmin46\arcsec$   \\
 R.A. (J2000.0)                   &    16$^h$13$^m$47$\fs7$           \\
 Dec. (J2000.0)                   &   +54$\degr22\arcmin15\arcsec$   \\
 Morphological type              &    dIrr/dSph       \\
				 &                               \\
 Angular diameter, $a_{26.5}$        &      1$\farcm1$          \\
 Axial ratio                     &      0.59                \\
 Extinction,  $A_V/A_I$              &    $0\fm03/0\fm02$             \\
 Total magnitude, $V_T$            &     $15\fm9\pm0\fm2$             \\
 $(V-I)_T$                         &     0.88$\pm$0.11              \\
				   &                                \\
 Central surface brightness, $\Sigma_{0,V}$&    $24\fm0\pm0\fm2/\sq\arcsec$ \\
 $I_{\rm TRGB}$                          &     $22\fm32\pm0\fm10$    \\
 $(m-M)_0$                         &    $26\fm35\pm0\fm14$       \\
 Distance,  $D_{MW}$                 &     1.86 Mpc           \\
 Distance,  $D_{LG}$                 &     1.79 Mpc            \\
				     &                         \\
 Linear diameter, $A_{26.5}$         &     0.54 Kpc             \\
 Absolute magnitude, $M_V$         &    $-$10.48      \\
 $(V - I)_{0,-3.5}$                  &    1.26$\pm$0.07   \\
  $\logz$                        &    $-$2.1$\pm$0.3 dex   \\
 Heliocentric velocity,  $V_h$     &   $-$135$\pm$2 km/s   \\
				   &                       \\
 Corrected velocity,  $V_{LG}$       &      +72  km/s    \\
 HI flux                         &     2.20 Jy$\cdot$km/s   \\
 HI line width, $W_{50}$         &           14 km/s       \\
 M(HI)/L$_V$                     &        0.80 $M_{\sun}/L_{\sun}$     \\
 M(dyn)/L$_V$                    &        1.50 $M_{\sun}/L_{\sun}$       \\
\hline
\end{tabular}
\end{table}
\onecolumn
\begin{figure}[hbt]
\vspace{17cm}
\caption{$R$-band image of KKR 25 in Draco obtained with the 6m SAO telescope. The horizontal
 line corresponds to 10$\arcsec$. North is up, and East to the left.}
\end{figure}
\begin{figure}[hbt]
\vspace{17cm}
\caption{WF3 image of KKR 25 produced by combining the two 600s exposures taken
through the F606W and F814W filters. A globular cluster candidate is indicated by the arrow.}
\end{figure}
\begin{figure}[hbt]
\vbox{\includegraphics{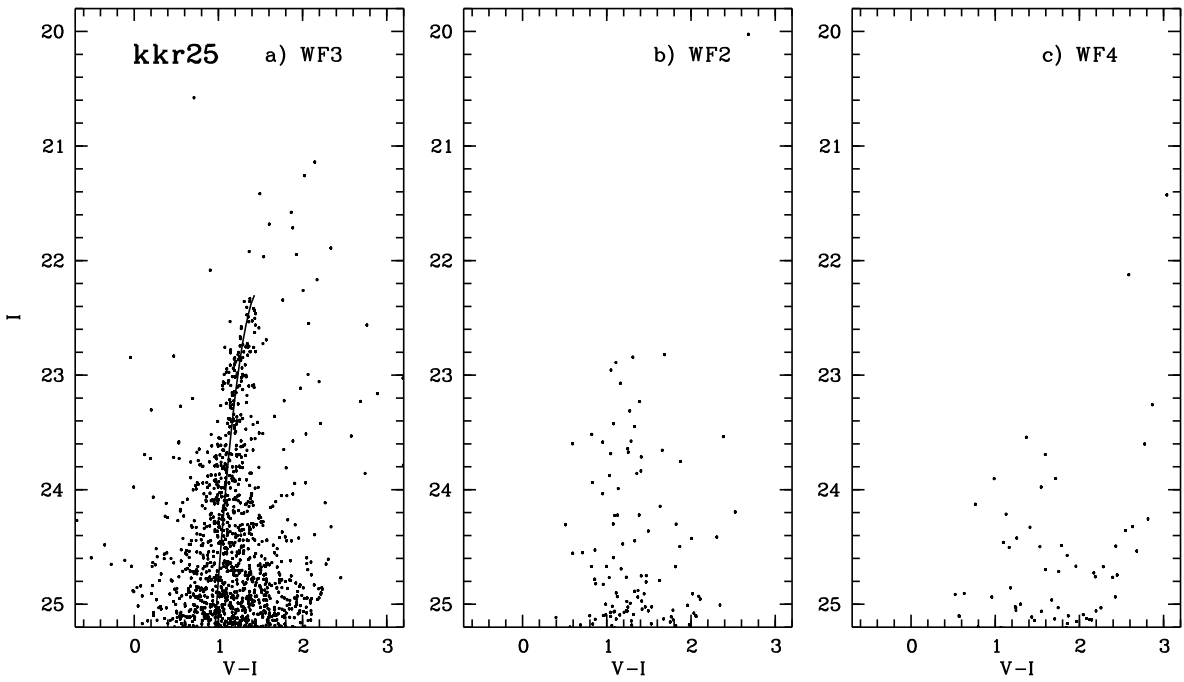}}\par
\vspace{20cm}
\caption{WFPC2 colour-magnitude diagram for KKR 25 in Draco. The left panel
shows stars in the WF3 chip, the middle and the right CMDs correspond to
the WF2 and WF4 chips. The solid line in the
left panel shows the red giant branch of the globular cluster
 M15 with metallicity of $-2.17$ dex.}
\end{figure}
\begin{figure}[hbt]
\vbox{\includegraphics{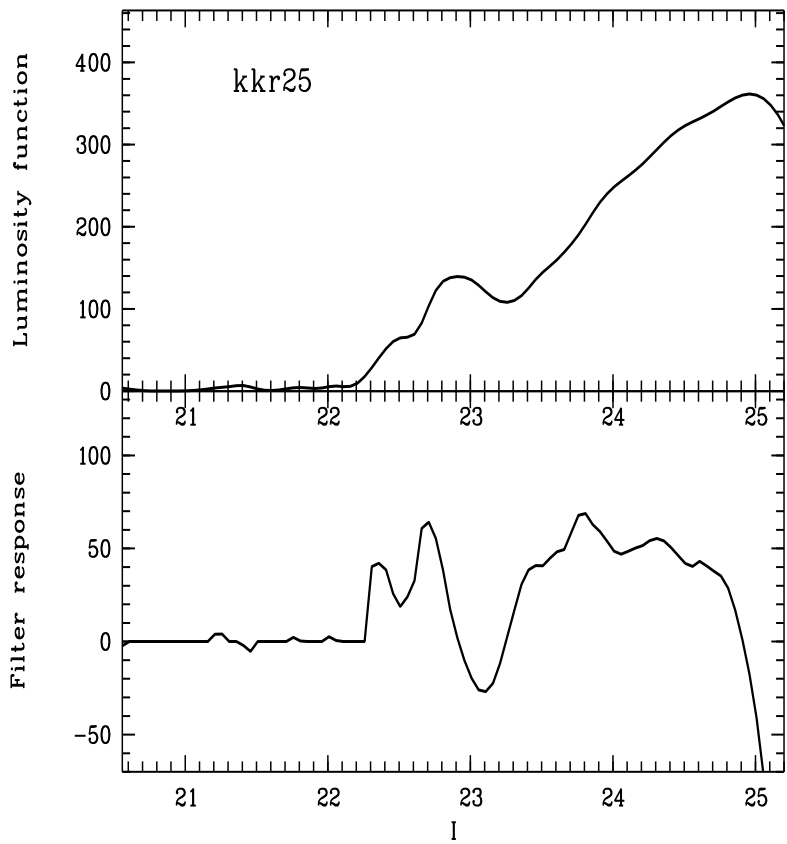}}\par
\vspace{20cm}
\caption{The Gaussian-smoothed I-band luminosity function restricted to red
stars with colors between $ 0.^{m}6 < (V - I) < 1.^{m}6 $ (top), and the
output of an edge-detection filter applied to the luminosity function (bottom)
for KKR 25.}
\end{figure}
\begin{figure}[hbt]
\vbox{\includegraphics{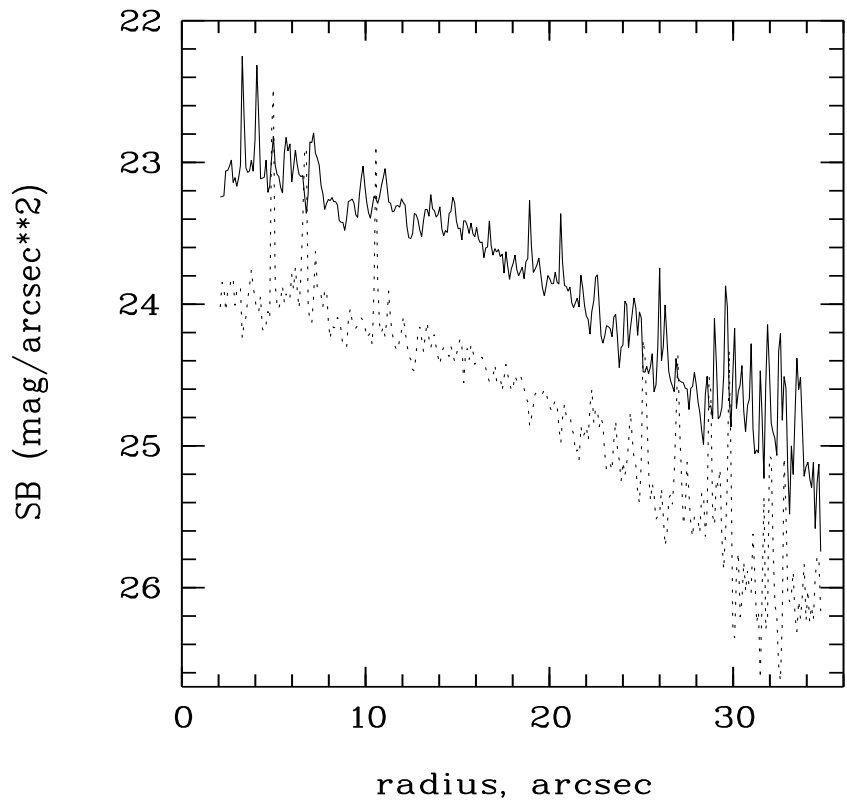}}\par
\vspace{20cm}
\caption{Radial distribution of surface brightness in KKR~25 in the $V$
(dashed) and the $I$ (solid line) bands.}

\end{figure}
\end{document}